\documentclass[twocolumn,prb,showpacs,floatfix]{revtex4}

\usepackage{amssymb}
\usepackage{amsmath}
\usepackage{epsfig}
\usepackage{graphics}
\usepackage[dvips]{color}

\begin{document}

\title{Cyclotron motion and magnetic focusing in semiconductor
quantum wells with spin-orbit coupling}

\author{John Schliemann}

\affiliation{Institute for Theoretical Physics, University of 
Regensburg, D-93040 Regensburg, Germany}

\date{\today}

\begin{abstract}
We investigate the ballistic motion of electrons in III-V 
semiconductor quantum wells with Rashba spin-orbit coupling 
in a perpendicular magnetic field. Taking into account the full
quantum dynamics of the problem, we explore the modifications of
classical cyclotron orbits due to spin-orbit interaction. 
As a result, for electron energies comparable with the
cyclotron energy the dynamics are particularly rich and not adequately
described by semiclassical approximations. Our study is 
complementary to previous semiclassical approaches concentrating on the
regime of weaker fields.
\end{abstract}

\maketitle

\section{Introduction}

The coupling between the orbital and the spin degrees of freedom of itinerant
carriers in semiconductors is a major direction of work
in today's spintronics research. An early key example is given by the
proposal of a spin field-effect transistor put forward by Datta and Das
already in 1990 \cite{Datta90}. The merit of this paradigmatic theoretical 
concept is that it allows, once realized, all-electrical control of electron 
spins in two-dimensional III-V semiconductor structures, avoiding any 
magnetic field. 
On the other hand, at about the same van Houten {\em et al.}
published a pioneering both theoretical and experimental study
of magnetic focusing of electrons in semiconductor quantum wells 
\cite{vanHouten89}. Here the 
control over the orbital degree of freedom of the carriers is achieved by
a perpendicular magnetic field of typically moderate strength, very analogously
to the classical cyclotron motion.
This effect has been demonstrated both for conduction-band electrons 
\cite{vanHouten89} and for valence-band holes \cite{Heremans92,Heremans93};
very recently cyclotron orbits in an electron focusing experiment were
directly imaged using scanning probe microscopy \cite{Aidala07}.

Other recent experimental studies have addressed the question whether
it is feasible to spatially separate (and, in turn, separately detect)
carriers in different spin-spilt
subbands of a quantum well via such transverse focusing techniques
\cite{Potok02,Rokhinson04,Rokhinson06}. In these investigations, the spin
splitting of subbands was either provided by a strong in-plane component of
the magnetic field \cite{Potok02} , or, more relevant for the present study, 
by specific contributions to spin-orbit coupling acting on the spin of 
carriers in semiconductor quantum wells
\cite{Rokhinson04,Rokhinson06,Dedigama06}. In the latter case, different 
initial spin states provide via spin-orbit interaction a separation of
carriers in real space, an effect which has already attracted also significant
theoretical interest
\cite{Goldoni91,Usaj04,Reynoso04,Valin-R06,Reynoso07a,Reynoso07b}.
In the present paper we provide a complementary theoretical study of 
cyclotron motion and magnetic focusing under the influence of spin-orbit
interaction using a fully quantum mechanical approach. 

A related phenomenon is the predicted {\em zitterbewegung} of carrier wave
packets in the presence of spin-orbit coupling \cite{Schliemann05,Schliemann06,Zawadzki05,Nikolic05,Shen05,Cserti06,Brusheim06,Rusin07,Schliemann07,Winkler07,Bernardes07,Zulicke07a,Zulicke07b}. {\em Zitterbewegung} of free electrons described by the
four-component Dirac equation was originally predicted by Schr\"odinger and
occurs for wave packets which contain solutions of the free Dirac equation of both positive and negative energy \cite{Schrodinger30}. In effective models for 
electrons and holes in semiconductors the intrinsic spin-dependent energy 
splitting due to spin-orbit coupling can lead to a similar oscillatory
{\em zitterbewegung}\cite{Schliemann05,Schliemann06,Cserti06,Winkler07}.
However, the latter effect is predicted to occur on time and length scales
being much more favorable for experimental detection compared to the situation
of free electrons where {\em zitterbewegung} has never been observed so far
\cite{Schliemann05,Schliemann06}. On the other hand, a unifying aspect of these
two phenomena is given by the fact that {\em zitterbewegung} of itinerant
band carriers in semiconductors occurs due to spin-orbit interaction which 
can be viewed as the nonrelativistic of the strong coupling between
spin and momentum being manifest in the Dirac equation.

Moreover, the interplay between spin-orbit coupling and cyclotron motion
in a perpendicular magnetic field was already studied theoretically in some
detail in Refs.~\cite{Winkler07,Zulicke07a}. Here the authors 
concentrate on semiclassical approximations, and on an analogy between
the Jaynes-Cummings model of atomic transitions in a radiation field and
the Rashba Hamiltonian \cite{Rashba60} 
in a perpendicular magnetic field, an aspect to be 
briefly reviewed below. In the present paper we report on numerical 
evaluations of the full quantum mechanical dynamics of a free electron
in a two-dimensional quantum well with spin-orbit interaction and a
perpendicular magnetic field, avoiding any further approximation.
As explained in the appendix, our approach is so far technically limited
to the Hilbert space of the first few ten lowest Landau levels. For
typical electron energies of a few meV, this restriction corresponds for
usual III-V semiconductor materials to magnetic fields of a few tesla.
Such fields are somewhat larger than those considered in circumstances of 
semiclassical approximations neglecting Landau quantization, and in this sense
our present study is complementary to those previous investigations.
For definiteness we will also concentrate
on spin-orbit coupling of the Rashba type, although
also other effective coupling terms can be considered. Finally we note that
a complementary theoretical study of conduction-band electrons being
subject to spin-orbit coupling and a homogeneous {\em in-plane electric field}
was given very recently in Ref.~\cite{Schliemann07}.

This paper is organized as follows. In section \ref{model} we summarize the
essential properties of the Rashba model in a perpendicular magnetic field.
We discuss the analogy to the Jaynes Cummings model of quantum optics, and we
describe in detail the initial states used for the numerical simulations
of time evolutions to be discussed in section \ref{results}. 
All further technical details can be found in the appendix. We close with
conclusions in section \ref{concl}.

\section{Model and approach}
\label{model}

We consider an electron in an n-doped quantum well being subject to 
Rashba spin-orbit coupling \cite{Rashba60}
and a homogeneous perpendicular magnetic field
coupling both to the orbital degrees of freedom as well as to the spin,
i.e. the single-particle Hamiltonian reads
\begin{equation}
{\cal H}=\frac{\vec \pi^{2}}{2m}
+\frac{\alpha}{\hbar}\left(\pi_{x}\sigma^{y}-\pi_{y}\sigma^{x}\right)
+\frac{1}{2}g\mu_{B}B\sigma^{z}.
\label{ham1}
\end{equation}
Here $m$ the the effective band mass, $\vec\pi=\vec p+e\vec A/c$ is the
two-component kinetic momentum with the canonical momentum $\vec p$
and the vector potential $\vec A$ generating the magnetic field
$\vec B$ along the growth direction of the quantum well chosen as the
$z$-axis, $\vec B=\nabla\times\vec A$. The effective Rashba spin-orbit coupling 
parameter is denoted by $\alpha$, g is the effective g-factor, $\mu_{B}$ the
Bohr magneton, and $\vec \sigma$ are the usual Pauli matrices.
Note that the Rashba Hamiltonian can be viewed as a 
momentum-dependent field coupling to the electron spin, an interpretation
we will use later on in the discussion of numerical results.
Moreover, in the following we will assume, without loss of generality, that
the product of the electron charge $(-e)=-|e|$ and the magnetic field
strength $B$ is always positive, $(-e)B>0$, i.e. $\vec B$ points along the
negative $z$-direction. 

\subsection{Spectrum and eigenstates}

Defining the usual bosonic operators
\begin{equation}
a=\frac{1}{\sqrt{2}}\frac{\ell}{\hbar}\left(\pi_{x}+i\pi_{y}\right)
\quad,\quad a^{+}=(a)^{+}
\end{equation}
fulfilling $[a,a^{+}]=1$ and  $\ell=\sqrt{\hbar c/|eB|}$
being the magnetic length, the Hamiltonian reads
\begin{eqnarray}
{\cal H} & = & \hbar\omega_{c}\left(a^{+}a+\frac{1}{2}\right)
+\frac{i}{\sqrt{2}}\frac{\alpha}{\ell}\left(a\sigma^{-}-a^{+}\sigma^{+}\right)\nonumber\\
& & +\frac{1}{2}g\mu_{B}B\sigma^{z}\,
\label{ham2}
\end{eqnarray}
where $\omega_{c}=|eB|/(mc)$ is the cyclotron frequency, and we have defined
$\sigma^{\pm}=\sigma^{x}\pm i\sigma^{y}$.
The operators $a$ and $a^{+}$ connect different Landau levels.
Note that the Hamiltonian
(including the spin-orbit part) can be expressed in terms of $a$ and $a^{+}$
only, no further orbital operators occur. 
Therefore its eigenstates have the same Landau level degeneracy 
as in the absence of spin-orbit coupling.

Fixing a certain intra-Landau-level
quantum number, we denote by 
$|n,\sigma\rangle=((a^{+})^{n}/\sqrt{n!})|0,\sigma\rangle$ 
a state in the $n$-th
Landau level with spin direction $\sigma\in\{\uparrow,\downarrow\}$.
Then $|0,\uparrow\rangle$ is an eigenstate
with energy $\varepsilon_{0}=(\hbar\omega_{c}+g\mu_{B}B)/2$, and 
all other eigenstates are of the form \cite{Rashba60,Schliemann03a}
\begin{equation}
|n,\pm\rangle=u^{\pm}_{n}|n,\uparrow\rangle+v^{\pm}_{n}|n-1,\downarrow\rangle
\label{speigen}
\end{equation}
with energy
\begin{equation}
\varepsilon_{n}^{\pm}=\hbar\omega_{c}n\pm
\sqrt{2n\frac{m\alpha^{2}}{\hbar^{2}}\hbar\omega_{c}+\frac{1}{4}
\left(\hbar\omega_{c}+g\mu_{B}B\right)^{2}}
\end{equation}
and the amplitudes
parametrizing the eigenstates read
\begin{eqnarray}
u^{\pm}_{n} & = & 
\left(\frac{1}{2}\pm\frac{\frac{1}{4}\left(\hbar\omega_{c}+g\mu_{B}B\right)}
{\sqrt{2n\frac{m\alpha^{2}}{\hbar^{2}}\hbar\omega_{c}+\frac{1}{4}
\left(\hbar\omega_{c}+g\mu_{B}B\right)^{2}}}\right)^{\frac{1}{2}}
\label{amp1}\\
v^{\pm}_{n} & = & \pm i\,{\rm sgn}(\alpha)\nonumber\\
 & & \cdot\left(\frac{1}{2}\mp
\frac{\frac{1}{4}\left(\hbar\omega_{c}+g\mu_{B}B\right)}
{\sqrt{2n\frac{m\alpha^{2}}{\hbar^{2}}\hbar\omega_{c}+\frac{1}{4}
\left(\hbar\omega_{c}+g\mu_{B}B\right)^{2}}}\right)^{\frac{1}{2}}\,.
\label{amp2}
\end{eqnarray}
Thus, the energy levels and eigenstates of the system are characterized by the 
interplay of three energy scales: The cyclotron energy $\varepsilon_{c}=\hbar\omega_{c}$, the
Zeeman energy $\varepsilon_{Z}=g\mu_{B}B$, and the Rashba energy $\varepsilon_{R}=m\alpha^{2}/ \hbar^{2}$.

\subsection{Analogy to the Jaynes-Cummings model}

As it was recognized recently in Ref.~\cite{Winkler07}, the Hamiltonian
(\ref{ham2}) is formally equivalent to the Jaynes-Cummings model for
atomic transitions in a radiation field. This model has been studied very
intensively in theoretical quantum optics, and the time evolution of
orbital and spin operators has been obtained in terms of analytical but
rather implicit expressions \cite{Ackerhalt75,Barnett97}. 
To explore this analogy it is useful to
separate the Hamiltonian into two commuting parts,
${\cal H}={\cal H}_{1}+{\cal H}_{2}$, with
\begin{eqnarray}
{\cal H}_{1} & = & \hbar\omega_c\left(a^{+}a+\frac{1+\sigma^{z}}{2}\right)\,,\\
{\cal H}_{2} & = & \frac{i}{\sqrt{2}}\frac{\alpha}{\ell}\left(a\sigma^{-}-a^{+}\sigma^{+}\right)
-\frac{\hbar\omega_{c}}{2}\left(1-\frac{gm}{2m_{0}}\right)\sigma^{z}\,,
\end{eqnarray}
where $m_{0}$ is the bare electron mass. Then the time evolution
of the position operators in the Heisenberg picture,
\begin{equation}
\vec r_{H}(t)=e^{iHt/ \hbar}\vec r(0)e^{-iHt/ \hbar}\,,
\end{equation}
can be written as
\cite{Winkler07,Ackerhalt75,Barnett97}
\begin{eqnarray}
 & & x_{H}(t)+iy_{H}(t)=x_{0}+iy_{0}\nonumber\\
 & & \qquad+\frac{ie^{-i(\omega_{c}+\omega_{+})t}}{\omega_{-}-\omega_{+}}
\left(\frac{\omega_{-}}{\omega_{c}}\frac{\pi_{x}+i\pi_{y}}{m}+i\frac{\alpha}{\hbar}\sigma^{+}\right)\nonumber\\
 & & \qquad-\frac{ie^{-i(\omega_{c}+\omega_{-})t}}{\omega_{-}-\omega_{+}}
\left(\frac{\omega_{+}}{\omega_{c}}\frac{\pi_{x}+i\pi_{y}}{m}+i\frac{\alpha}{\hbar}\sigma^{+}\right)\,,
\label{acker}
\end{eqnarray}
where the operator-valued frequencies $\omega_{\pm}$ are given by
\begin{equation}
\hbar\omega_{\pm}=-{\cal H}_{2}\pm\sqrt{2\frac{m\alpha^{2}}{\hbar^{2}}\hbar\omega_{c}+{\cal H}_{2}^{2}}
\end{equation}
and $x_{0}$, $y_{0}$ are the usual coordinates of the center of the classical
cyclotron orbit which commute with the Hamiltonian and are therefore
constant in time.

The result (\ref{acker}) is correct but still not very explicit. 
In particular, the
operator character of the quantities $\omega_{\pm}$ poses severe obstacles against
evaluating this expression for a given initial state. Therefore, in the 
present work we follow a different route towards the full quantum dynamics
by expanding the initial state of the system in terms of its eigenstates.

\subsection{Gauge and initial state}

In the very general considerations so far there was not any necessity to
specify the gauge of the vector potential $\vec A$. For the practical 
calculations to be described below, however, we shall work in the Landau gauge 
$\vec A=(0,Bx,0)$ where the spinless orbital eigenstates in the absence of 
spin-orbit coupling have the following well-known form
\begin{eqnarray}
\langle\vec r|n,k\rangle  & = & \frac{i^{n}}{\sqrt{n!2^{n}\ell\sqrt{\pi}}}
H_{n}\left(\frac{x-k\ell^{2}}{\ell}\right)\nonumber\\
 & & \times\exp\left(-\frac{1}{2\ell^{2}}\left(x-k\ell^{2}\right)\right)
\frac{e^{iky}}{\sqrt{2\pi}}
\label{basis1}
\end{eqnarray}
labelled by a wave number $k$ corresponding to translational invariance in the
$y$-direction, or, equivalently, by a guiding center coordinate $k\ell^{2}$
for the $x$-direction. $H_{n}(x)$ are the usual Hermite polynomials, and the
phases of the above wave functions have been adjusted to fulfill
$a|n,k\rangle=\sqrt{n}|n-1,k\rangle$.

In what follows we will be interested in the quantum dynamics of an initial
state $|\psi\rangle$ being a direct product of an orbital and a spin state,
\begin{equation}
|\psi\rangle=|\phi\rangle\left(
\begin{array}{c}
\kappa \\
\lambda
\end{array}
\right)\,,
\label{init}
\end{equation}
where the spinor components are related to the usual polar angles
$\vartheta$, $\varphi$ of the initial spin direction
via $\kappa=\exp(-i\varphi/2)\cos(\vartheta/2)$, $\lambda=\exp(i\varphi/2)\sin(\vartheta/2)$. As a generic initial 
orbital state we consider
\begin{equation}
\langle\vec r|\phi\rangle=\frac{1}{\sqrt{\pi}d}e^{-\frac{r^{2}}{2d^{2}}+ik_{0}y}\,,
\label{initorb}
\end{equation}
i.e. a normalized Gaussian wave packet of spatial width $d$ and initial
momentum $\hbar k_{0}$ along the $y$-axis, i.e. the direction of translational 
invariance of the Hamiltonian. The initial position of the particle is at
the origin, $\langle\psi|\vec r|\psi\rangle=0$.

The energy of the above initial state can be expressed as
\begin{eqnarray}
\langle\psi|{\cal H}|\psi\rangle & = & \frac{1}{2}\hbar\omega_{c}\left(\frac{\ell^{2}}{d^{2}}+\frac{d^{2}}{2\ell^{2}}
+k_{0}^{2}\ell^{2}\right)\nonumber\\
 & & -\sqrt{\frac{m\alpha^{2}}{\hbar^{2}}\hbar\omega_{c}}k_{0}\ell\left(\bar\kappa\lambda+\bar\lambda\kappa\right)\nonumber\\
 & & + \frac{1}{2}g\mu_{B}B\left(\bar\kappa\kappa-\bar\lambda\lambda\right)
\end{eqnarray}
where the bar denotes complex conjugation. Conversely, for a wave packet
of the above form with fixed energy $E$ and given width and spin state, the
initial momentum reads
\begin{eqnarray}
k_{0}\ell & = & \sqrt{\frac{\frac{m\alpha^{2}}{\hbar^{2}}}{\hbar\omega_{c}}}\left(\bar\kappa\lambda+\bar\lambda\kappa\right)
\nonumber\\
 & \pm & \Biggl[\frac{2E}{\hbar\omega_{c}}-\left(\frac{\ell^{2}}{d^{2}}+\frac{d^{2}}{2\ell^{2}}
+g\mu_{B}B\left(\bar\kappa\kappa-\bar\lambda\lambda\right)\right)\nonumber\\
 & & +\frac{\frac{m\alpha^{2}}{\hbar^{2}}}{\hbar\omega_{c}}\left(\bar\kappa\lambda+\bar\lambda\kappa\right)^{2}\Biggr]^{1/2}\,.
\label{wverg}
\end{eqnarray}

\subsection{Time evolution}

A conceptually straightforward way to evaluate time-dependent expectation 
values is to expand the initial state in terms of the eigenstates of the
above system and use the matrix elements of the desired operator in this
eigenbasis. For instance, for the kinetic-momentum operator, this approach 
formally reads
\begin{eqnarray}
 & & \langle\psi|\vec \pi_{H}(t)|\psi\rangle=\sum_{n_{1},n_{2}=0}^{\infty}\sum_{\mu_{1},\mu_{2}}\int_{-\infty}^{\infty}dk
\Biggl[\langle\psi|n_{1},k,\mu_{1}\rangle\nonumber\\
 & & \qquad\times\langle n_{1},k,\mu_{1}|\vec \pi|n_{2},k,\mu_{2}\rangle\langle n_{2},k,\mu_{2}|\psi\rangle\nonumber\\
 & & \qquad\times
\exp\left(\frac{i}{\hbar}\left(\varepsilon_{n_{1}}^{\mu_{1}}-\varepsilon_{n_{2}}^{\mu_{2}}\right)t\right)\Biggr]\,,
\label{timeevol}
\end{eqnarray}
where we have already anticipated that these operators are diagonal in 
the intra-Landau-level quantum number $k$; the same holds for the  
spin operators $\vec \sigma_{H}(t)$. The summation over $\mu_{i}$, 
$i\in\{1,2\}$ runs over $\mu_{i}=\pm$ for $n_{i}>0$ and $\mu_{i}=\uparrow$ for $n_{i}=0$.
From the expectation values of the position operators can be obtained from 
those of the momenta via
\begin{eqnarray}
x & = & x_{0}+\frac{c}{eB}\pi_{y}\,,\\
y & = & y_{0}-\frac{c}{eB}\pi_{x}\,,
\end{eqnarray}
where the expectation values of the constant centers of the
classical cyclotron motion are given by $\langle\psi|x_{0}|\psi\rangle=k_{0}\ell^{2}$, $\langle\psi|y_{0}|\psi\rangle=0$. 
As explained in detail in the appendix, the integration over the
wave numbers $k$ can be performed separately and serves as an input for 
the numerical evaluation of the remaining sums. For any further technical 
details, we refer the reader to the appendix.

\section{Results}
\label{results}

We now present the results of numerical simulations of the
time evolution of expectation values described in Eq.~(\ref{timeevol}).
All relevant technical details can be found in the appendix. In all simulation
we assume the Rashba coefficient to be positive, $\alpha>0$.

\subsection{Cyclotron motion}
\label{cyclomotion}

\begin{center}
\begin{figure}
\epsfig{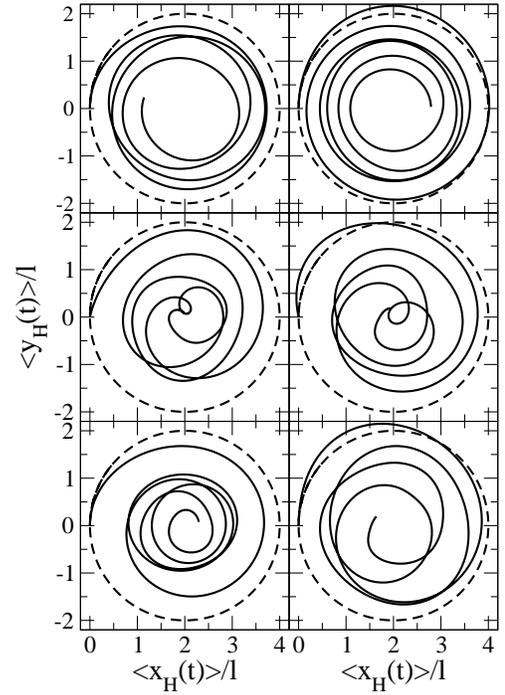}
\caption{Orbital dynamics of a wave packet of initial width $d=1.0\ell$ and
group wave number $k_{0}=2.0/ \ell$ for various initial spin states.
The Rashba energy is $\varepsilon_{R}=0.2\hbar\omega_{c}$ while the Zeeman energy is put to zero.
In the left (right) top panel, the spin points initially along the
positive (negative) $x$-direction. The middle and bottom panels
show the corresponding data for the $y$- and $z$-direction, respectively.
The simulation time is always $t=30/ \omega_{c}$. The stricly circular motion
(dotted lines) 
with radius $k_{0}\ell^{2}$ occurring in the absence of spin-orbit coupling is
always shown as a guide to the eye.}
\label{fig1}
\end{figure}
\end{center}

Let us first investigate the influence of spin-orbit coupling on the
cyclotron motion in general.
Fig.~\ref{fig1} shows the the particle orbit evaluated in terms
of the expectation values $\langle\psi|\vec r_{H}(t)|\psi\rangle=:\langle\vec r_{H}(t)\rangle$
of a wave packet of initial width $d=1.0\ell$ and
group wave number $k_{0}=2.0/ \ell$ for various initial spin states.
The Rashba energy is $\varepsilon_{R}=0.2\hbar\omega_{c}$ while the Zeeman energy is, for simplicity,
put to zero here.
In the left (right) top panel, the spin points initially along the
positive (negative) $x$-direction. The middle and bottom panels
show the corresponding data for the $y$- and $z$-direction, respectively.
The total simulation time is always $t=30/ \omega_{c}$. The strictly circular motion
(dotted lines) 
with radius $k_{0}\ell^{2}$ occurring in the absence of spin-orbit coupling is
shown in all graphs as a guide to the eye. The magnetic length $\ell$
can conveniently converted into practical units via 
$\ell=257 \AA / \sqrt{B/ {\rm Tesla}}$.

All six graphs have the appearance of a more or less distorted spiral.
The {\em prima vista} most regular motion is found in the two top panels where
the initial spin direction is collinear with in initial direction 
of the momentum-dependent coupling to the spin described by the Rashba 
Hamiltonian. This situation was investigated very recently in 
Ref.~\cite{Zulicke07a} in the framework of several schemes of
semiclassical approximations. In one of these approaches the
spin is assumed to follow in an adiabatic fashion the momentum-dependent
field coupling to it, where both quantities are taken to be classical
variables. We will discuss below to what extend this approximation
leads to useful results in the parameter regime considered here where
the energy of the initial wave packet is comparable with the 
cyclotron energy.

\begin{center}
\begin{figure}
\epsfig{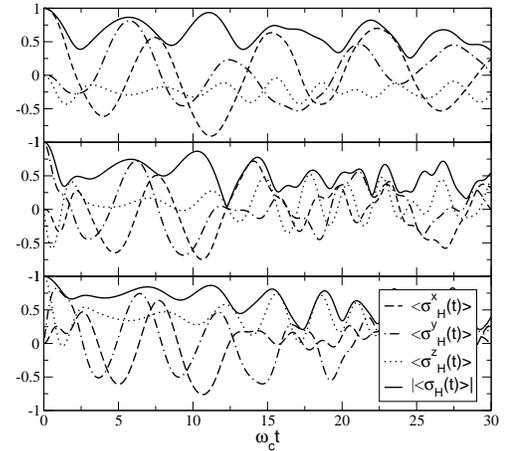}

\caption{Spin dynamics as expressed in terms of the
time-dependent expectation values corresponding to the left column of
Fig.~\ref{fig1}. The solid lines show the quantity 
$|\langle\vec\sigma_{H}(t)\rangle|$ which is a measure of entanglement between the spin and the 
orbital degrees of freedom.}
\label{fig2}
\end{figure}
\end{center}

In Fig.~\ref{fig2} we have plotted the corresponding 
spin dynamics expressed in terms of the
time-dependent expectation values $\langle\vec\sigma_{H}(t)\rangle$. The initial conditions
in the three panels are the same as in the left column of Fig.~\ref{fig1}.
The solid lines show the modulus of the vector $|\langle\vec\sigma_{H}(t)\rangle|$ of the 
time-dependent expectation values of spin components. This quantity can be
used as a measure of entanglement between the electron spin and its orbital 
degrees of freedom \cite{Schliemann02,Bennett96}. 
In fact, when tracing out the real-space degrees of 
freedom, the time-dependent reduced density matrix of the spin reads
\begin{eqnarray}
\rho_{spin}(t) & = & {\rm tr}_{orb}\left[e^{-\frac{i}{\hbar}{\cal H}t}|\psi\rangle\langle\psi|e^{\frac{i}{\hbar}{\cal H}t}\right]\nonumber\\
 & = & \frac{1}{2}\left(
\begin{array}{cc}
1+\langle\sigma^{z}_{H}(t)\rangle & \langle\sigma^{+}_{H}(t)\rangle \\
\langle\sigma^{-}_{H}(t)\rangle & 1-\langle\sigma^{z}_{H}(t)\rangle 
\end{array}
\right)
\end{eqnarray}
with eigenvalues $\lambda_{\pm}(t)=(1\pm|\langle\vec\sigma_{H}(t)\rangle|)/2$.
Thus, a modulus of $|\langle\vec\sigma_{H}(t)\rangle|=1$ (as present in the intitial condition
at $t=0$) corresponds to a direct product of spin and orbital state
with a reduced spin density matrix of rank $1$, while a vanishing
modulus $|\langle\vec\sigma_{H}(t)\rangle|=0$ indicates maximal entanglement between
spin and orbital degrees of freedom, and the reduced spin density matrix is
proportional to the unit matrix \cite{Schliemann02,Bennett96}.
As seen in Fig.~\ref{fig2}, the modulus $|\langle\vec\sigma_{H}(t)\rangle|$ is generically
clearly smaller
than unity signalling between spin and real-space coordinates, an effect
certainly beyond semiclassical approximations.

Let us now come back to the investigation of
the adiabatic semiclassical approximation employed in
Ref.~\cite{Zulicke07a}. 
Here both spin and particle momentum are treated as classical variables, and 
the projection of the spin on the instantaneous direction of the 
momentum-dependent field is assumed to be constant. Intuitively, this 
assumption corresponds to strong spin-orbit coupling \cite{Zulicke07a}
as it is the case for a Rashba energy of $\varepsilon_{R}=0.2\hbar\omega_{c}$ studied above.
To investigate the validity
of this adiabatic approximation, we introduce
\begin{equation}
Q_{1}(t):=\frac{\langle(\pi_{x})_{H}(t)\rangle\langle\sigma^{y}_{H}(t)\rangle-\langle(\pi_{y})_{H}(t)\rangle\langle\sigma^{x}_{H}(t)\rangle}{|\langle\vec\pi_{H}(t)\rangle|}
\end{equation}
and 
\begin{equation}
Q_{2}(t):=\frac{\langle(\pi_{x})_{H}(t)\rangle\langle\sigma^{y}_{H}(t)\rangle-\langle(\pi_{y})_{H}(t)\rangle\langle\sigma^{x}_{H}(t)\rangle}
{|\langle\vec\pi_{H}(t)\rangle||\langle\vec\sigma_{H}(t)\rangle|}
\end{equation}
The first quantity is 
the projection of the vector $\langle\vec\sigma_{H}(t)\rangle$ onto the direction of the 
momentum-dependent field
evaluated in terms of $\langle\vec\pi_{H}(t)\rangle$, whereas in $Q_{2}$ we have additional 
divided by $|\langle\vec\sigma_{H}(t)\rangle|$ in order to eliminate the effects of
entanglement discussed above.
For the adiabatic-semiclassical 
approximation to be valid, $Q_{1}$ and $Q_{2}$ should be reasonably constant in 
time. 

\begin{center}
\begin{figure}
\epsfig{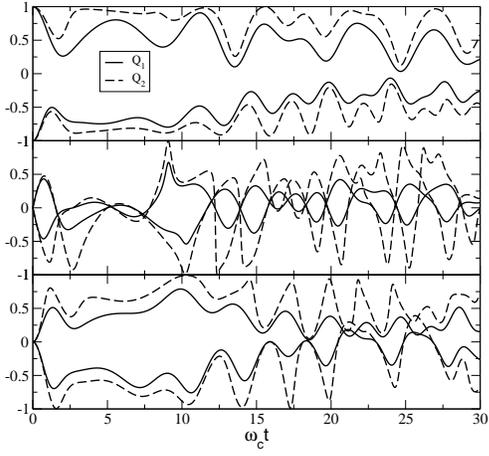}

\caption{The quantities $Q_{1}$ (solid lines) and $Q_{2}$ (dashed lines)
defined in the text as a function of 
time for the same system parameters as in Fig.~\ref{fig1}.
In the top panel, the spin points initially along the positive (negative)
$x$-direction with $Q_{1}(0)=Q_{2}(0)=+1$ ($Q_{1}(0)=Q_{2}(0)=-1$).
The middle and bottom panel show the analogous data with the
spin initially aligned along the $y$- and $z$-axis, respectively.
Here we have always $Q_{1}(0)=Q_{2}(0)=0$.}
\label{fig3}
\end{figure}
\end{center}

\begin{center}
\begin{figure}
\epsfig{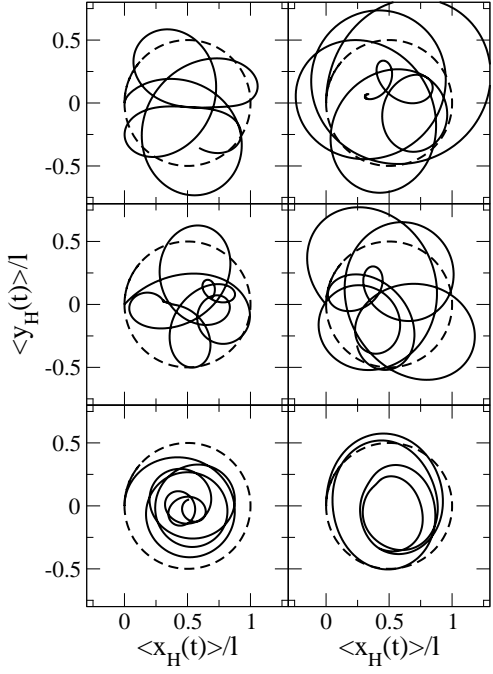}

\caption{Orbital dynamics for the same system as in Fig.~\ref{fig1} but
with a smaller initial group wave number of only $k_{0}=0.5/ \ell$.
Again, the stricly circular motion
occurring in the absence of spin-orbit coupling is
always shown as a guide to the eye.}
\label{fig4}
\end{figure}
\end{center}

Fig.~\ref{fig3} shows the time dependence of $Q_{1}$ and $Q_{2}$ 
for the same system parameters as in Fig.~\ref{fig1}. In particular
$\langle\vec\pi_{H}(0)\rangle$ points along the positive $y$-direction which means that
the momentum-dependent field coupling to the spin is initially 
in the $x$-direction in spin space.
In the top panel, the spin points initially along the positive (negative)
$x$-direction with $Q_{1}(0)=Q_{2}(0)=+1$ ($Q_{1}(0)=Q_{2}(0)=-1$).
The middle and bottom panels show the analogous data with the
spin initially aligned along the $y$- and $z$-axis, respectively.
Here we have always $Q_{1}(0)=Q_{2}(0)=0$. As seen in the figure,
$Q_{1}(t)$ and $Q_{2}(t)$ significantly deviate from a constant value even in the
case where the spin is initially fully aligned with the momentum-dependent
field (top panel). From these observations we conclude that this adiabatic
semiclassical approximation is rather problematic in the
parameter regime studied here where the total energy of the electron
wave packet is of the order of the cyclotron energy.  
In fact, the behavior of the system is only rather poorly represented by
introducing two different cyclotron radii corresponding to two
spin directions as suggested in Ref.~\cite{Zulicke07a}. On the contrary, the 
dynamics are much richer and show trajectories reminiscent of chaotic
behavior. The latter observation becomes even more significant
at smaller initial group wave number $k_{0}$ as one can see in Fig.~\ref{fig4}
where we have plotted trajectories analogous to those in Fig.~\ref{fig1}
but with a shorter initial group wave vector of only $k_{0}=0.5/ \ell$.
For lower cyclotron energies, however, we expect the semiclassical
approximation of Ref.~\cite{Zulicke07a} to be significantly better
fulfilled than in the regime studied here. The smaller the cyclotron energy 
compared to the total energy of the wave packet, the larger the
number of Landau levels to be included in the numerical simulation.
As explained in the appendix, such simulations require the
precise numerical evaluation of high-order Hermite polynomials, a task which
technically limits our approach to the regime where the cyclotron energy
is comparable with the energy of the initial wave packet.
In this sense, our study is 
complementary to previous semiclassical approaches concentrating on the
regime of weaker fields.

\begin{center}
\begin{figure}
\epsfig{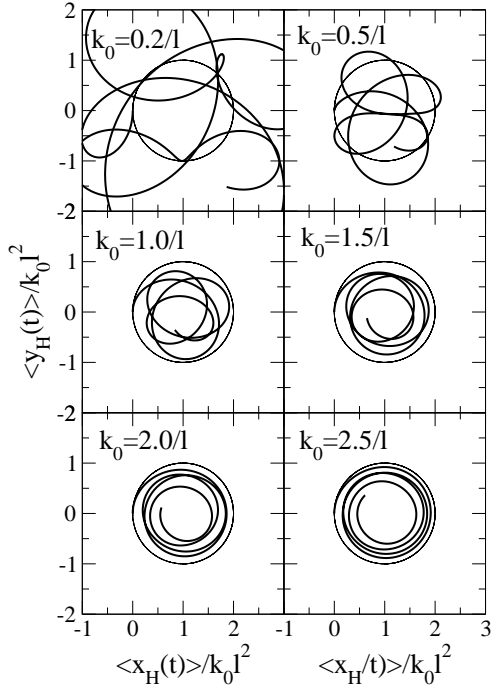}
\caption{Orbital dynamics for the same system as in Fig.~\ref{fig1} 
for various values of the  initial group wave number 
$k_{0}$ and the spin initially always pointing along the positive $x$-axis.
For a better comparison the the components of $\langle\vec r\rangle$ are given
in units of $k_{0}\ell^{2}$.}
\label{fig5}
\end{figure}
\end{center}

Finally Fig.~\ref{fig5} shows
the orbital dynamics for again the same system as in 
Fig.~\ref{fig1} for various values of the  initial group wave number 
$k_{0}$ and the spin initially always pointing along the positive $x$-axis.
For a better comparison the the components of $\langle\vec r\rangle$ are given
in units of $k_{0}\ell^{2}$. Clearly, the dynamics become more ``regular'' (or less
``chaotic'') the larger the initial group wave number $k_{0}$.

In the above investigations we have concentrated on the 
position operator $\vec r_{H}(t)$ to describe the
time evolution of the initial state chosen as a Gaussian wave packet.
Regarding the width of this wave packet 
(as opposed to its center $\langle\vec r_{H}(t)\rangle$) let us consider the
case of vanishing spin-orbit coupling. Here the components of the
position operator are straightforwardly obtained as
\begin{eqnarray}
x_{H}(t) & = & x(0)+\frac{\pi_{x}(0)}{m\omega_{c}}
\sin\left(\omega_{c}t\right)\nonumber\\
 & & +\frac{\pi_{y}(0)}{m\omega_{c}}
\left(1-\cos\left(\omega_{c}t\right)\right)\,,\\
y_{H}(t) & = & y(0)+\frac{\pi_{y}(0)}{m\omega_{c}}
\sin\left(\omega_{c}t\right)\nonumber\\
 & & -\frac{\pi_{x}(0)}{m\omega_{c}}
\left(1-\cos\left(\omega_{c}t\right)\right)\\
\end{eqnarray}
and are completely analogous to the classical cyclotron motion.
For an initial state Gaussian wave packet given in Eq.~(\ref{initorb})
the time dependent width reads
\begin{eqnarray}
 & & \langle x^{2}_{H}(t)\rangle-\langle x_{H}(t)\rangle^{2}
+\langle y^{2}_{H}(t)\rangle-\langle y_{H}(t)\rangle^{2}\nonumber\\
 & & \quad=d^{2}+\left(2\frac{\ell^{4}}{d^{2}}-d^{2}\right)
\left(1-\cos\left(\omega_{c}t\right)\right)\,.
\end{eqnarray}
Thus, differently from the dispersive dynamics of a wave packet in the
absence of a magnetic field, the width does not increase to infinity
but remains bounded and rather oscillates with the cyclotron frequency,
similarly to the time evolution of a coherent state in a harmonic oscillator.
In the presence of spin-orbit coupling we expect the time evolution of
the width of the initial state to be more complex but still essentially
bounded. 

\subsection{Magnetic focusing}

\begin{center}
\begin{figure}
\epsfig{file=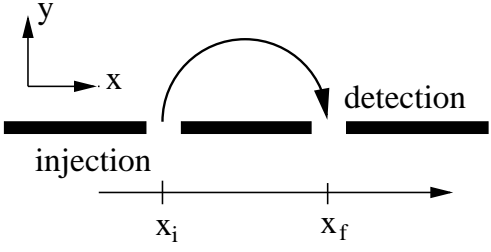,width=0.36\textwidth}
\caption{Schematic sketch of a magnetic focusing experiment.}
\label{fig6}
\end{figure}
\end{center}

Let us now turn to the issue of magnetic focusing under the influence of
spin-orbit coupling. A magnetic focusing experiment is conceptually very simple
and sketched in Fig.\ref{fig6}: Electrons enter a quantum well
at a location $x=x_{i}$, follow a ballistic cyclotron orbit, and
impinge again on the boundary of the system at a location 
$x=x_{f}$. In the absence of spin-orbit coupling,  this difference in 
coordinate depends, just as in the
classical case, only on the applied magnetic field and the initial
group wave number $k_{0}$, $x_{f}-x_{i}=2k_{0}\ell^{2}$. Thus, using an appropriately
located detection contact one can study electron transport as a function of 
these two quantities. 

\begin{center}
\begin{figure}
\epsfig{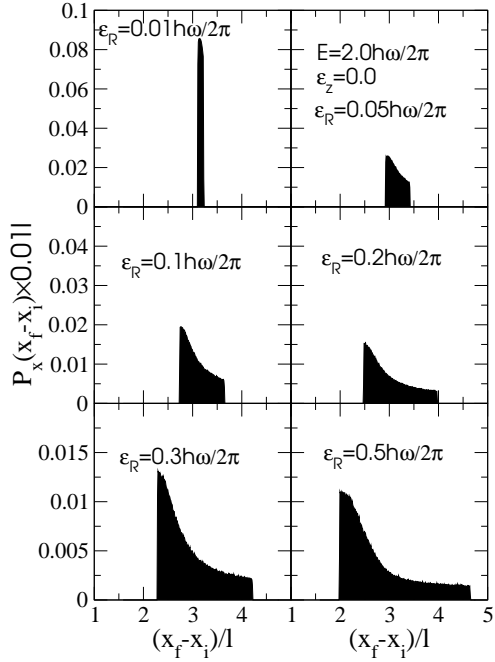}
\caption{The probability density $P_{x}(x_{f}-x_{i})$ for a wave packet of 
width $d=\ell$, total and energy $E=2.0\hbar\omega$ 
at zero Zeeman coupling and different Rashba energies. 
The data is averaged over
500000 randomly chosen initial spin states each. Note the different
scale of the $y$-axis in the top, middle, and bottom panels.}
\label{fig7}
\end{figure}
\end{center}

\begin{center}
\begin{figure}
\epsfig{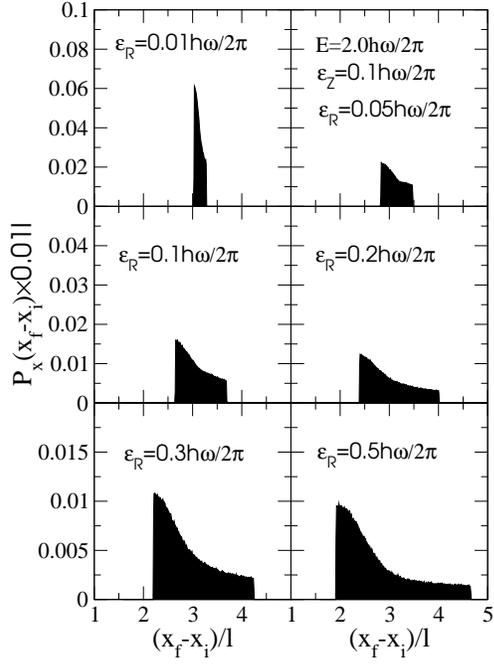}
\caption{The probability density $P_{x}(x_{f}-x_{i})$ for the same situation as in 
Fig.\ref{fig7} but with a Zeeman energy of $\varepsilon_{z}=0.1\hbar\omega$.}
\label{fig8}
\end{figure}
\end{center}

In a typical experiment, 
however, rather the energy $E$ of electrons 
(defined by the Fermi energy of the injecting lead)
not their momentum is fixed 
with both quantities being connected via Eq.~(\ref{wverg}).
Therefore, the wave number $k_{0}$ of an injected electron will
in general depend on its spin state;
only in the absence of both Zeeman coupling and spin-orbit interaction
$k_{0}$ is independent of the electron spin.
In turn, for random initial spin directions, $x_{f}-x_{i}$ will be distributed 
according to some probability density $P_{x}(x_{f}-x_{i})$. Fig.\ref{fig7} shows
a numerical evaluation of $P_{x}(x_{f}-x_{i})$ for a wave packet of 
width $d=\ell$, total and energy $E=2.0\hbar\omega$ 
at zero Zeeman coupling and different Rashba energies.
The initial angular coordinates
$\cos\vartheta$ and $\varphi$ determining the complex amplitudes
$\kappa$ and $\lambda$ in the initial state (\ref{init}) were chosen at random
from uniform distributions in the interval $[-1,1]$ and $[0,2\pi]$, respectively.
The data is averaged over
500000 randomly chosen initial spin states each. At a small Rashba
energy of only $\varepsilon_{R}=m(\alpha/ \hbar)^{2}=0.01\hbar\omega$ (top left panel)
the dynamics depend only very weakly
on spin and $P_{x}(x_{f}-x_{i})$ is strongly peaked around $x_{f}-x_{i}=2k_{0}\ell^{2}\approx3.16$, the 
classical cyclotron diameter expected in the absence of spin-orbit coupling.
With increasing Rashba energy, this peak undergoes a broadening with
the maximum of the probability density being located at smaller
values of $x_{f}-x_{i}$. Fig.~\ref{fig8} shows the same type of data at a
Zeeman energy of $\varepsilon_{Z}=0.1\hbar\omega$ which does not lead to any qualitative difference.

In summary, the initial narrow peak 
of the probability density $P_{x}(x_{f}-x_{i})$ at small spin-orbit coupling broadens
with increasing Rashba energy and develops an non-trivial structure
in terms of a maximum at small arguments with a broad shoulder reaching to
higher values. 
The structures seen in Figs.~\ref{fig7},\ref{fig8} appear to be
somewhat different to the results of Ref.~\cite{Usaj04} where a splitting
of the conductance peak as a function of magnetic field was found for
increasing Rashba coupling. These two peaks can be related 
to two different effective cyclotron radii corresponding to two
initial spin states with respect to a quantization axis being perpendicular
to the initial momentum and the magnetic field, an observation  similar
to the ``strong coupling'' semiclassical scenario of Ref.~\cite{Zulicke07a}.
Thus, in the light of these investigations one
could also expect a double-peak structure to develop in 
the probability density $P_{x}(x_{f}-x_{i})$. However, the investigations of
Ref.~\cite{Usaj04} work at clearly higher electron energies compared to
the cyclotron energy, and Landau quantization is explicitly neglected.
This is in contrast to the present study which works at larger
cyclotron energies taking into account the full quantum dynamics of the
problem.

\section{Conclusions and Outlook}
\label {concl}

We have studied the ballistic motion of electrons in III-V 
semiconductor quantum wells with Rashba spin-orbit coupling 
and a perpendicular magnetic field. Differently from previous
investigations, our numerical approach takes into account
the full quantum mechanics of the problem and is technically
limited to situations where the cyclotron energy is of the same order
as the energy of the initial electron wave packet. For typical
experimental parameters, this restriction corresponds to magnetic
fields of a few tesla. Such fields are larger than those considered 
previously in circumstances of 
semiclassics neglecting Landau quantization, and in this sense
our present study is complementary to such semiclassical approaches.
As a result, in the parameter regime considered here
the electron dynamics are particularly rich and not adequately
described by semiclassical approximations. An interesting issue
for futher investigations here includes the question whether the
seemingly ``chaotic'' trajectories shown in section \ref{cyclomotion}
are truely ergodic. Moreover, it is tempting to attribute
the irregularity of these trajectories to the {\em zitterbewegung}
predicted previously for electron motion in two-dimensional electron
gases without magnetic fields \cite{Schliemann05,Schliemann06,Zawadzki05,Nikolic05,Shen05,Cserti06,Brusheim06,Rusin07,Schliemann07,Winkler07,Bernardes07,Zulicke07a,Zulicke07b}.
What both phenomena have indeed in common is the fact
that they are the result of spin-orbit coupling, and the 
irregular motion of electrons in a perpendicular magnetic field
is the consequence of the non-equidistant spectrum of Landau levels 
induced by spin-orbit interaction.

In this study, we have concentrated on spin-orbit coupling of the
Rashba type. However, the situation of linear Dresselhaus coupling
can be treated analogously \cite{Schliemann03a}
as it only couples, like the Rashba term,  pairs of neighboring Landau 
levels with opposite spin. 
If both types of spin-orbit coupling terms are present, 
all Landau levels are coupled, and the single-particle
Hamiltonian cannot be diagonalized analytically anymore. In this case
the eigensystem of the Hamiltonian needs to be computed numerically,
or appropriate approximations have to be employed
\cite{Valin-R06}. It is an interesting question whether the inclusion
of both kinds of couplings leads to qualitativly new observations.
A particular situation is reached 
if both terms occur with the same magnitude, 
where, for zero Zeeman coupling, a new conserved spin operator arises
\cite{Schliemann03b}.

Further possible extensions of the present work include
the study of valence-band holes (as opposed to conduction-band electrons)
with an effective spin-orbit coupling being trilinear in the momentum,
and electron or hole dynamics under the influence of an additional
in-plane electric field.

\acknowledgments{I thank S.~Q. Shen  and U. Z\"ulicke for useful discussions. 
This work was supported by DFG via SFB 689  
``Spin Phenomena in reduced Dimensions''.}

\appendix

\section{Technical details}

\subsection{Overlap with basis states}

The overlap of the initial orbital state (\ref{initorb}) and the 
basis states (\ref{basis1}) of the usual Landau levels can be
expressed as
\begin{eqnarray}
\langle n,k|\phi\rangle & =& (-i)^{n}\sqrt{\frac{2}{n!2^{n}}}\sqrt{\frac{d}{\sqrt{\pi}}}
\frac{\sqrt{\frac{d}{\ell}}}{\sqrt{1+\frac{d^{2}}{\ell^{2}}}}
\left(\frac{1-\frac{d^{2}}{\ell^{2}}}{1+\frac{d^{2}}{\ell^{2}}}\right)^{n/2}\nonumber\\
 & & \times\exp\left(-\frac{1}{2}\frac{k_{0}^{2}\ell^{2}}{1+\frac{d^{2}}{\ell^{2}}}
-\frac{d^{2}}{2}\left(k-k_{0}\right)^{2}\right)\nonumber\\
 & & \times H_{n}\left(-\frac{k_{0}\ell}{\sqrt{\left(1-\frac{d^{2}}{\ell^{2}}\right)
\left(1+\frac{d^{2}}{\ell^{2}}\right)}}\right)
\label{ovl1}\\
 & = & (-i)^{n}\sqrt{\frac{2}{n!2^{n}}}\sqrt{\frac{d}{\sqrt{\pi}}}
\frac{\sqrt{\frac{d}{\ell}}}{\sqrt{1+\frac{d^{2}}{\ell^{2}}}}\nonumber\\
 & & \times\exp\left(-\frac{1}{2}\frac{k_{0}^{2}\ell^{2}}{1+\frac{d^{2}}{\ell^{2}}}
-\frac{d^{2}}{2}\left(k-k_{0}\right)^{2}\right)\nonumber\\
 & & \times\sum_{p=0}^{\left[\frac{n}{2}\right]}\Biggl[\frac{(-1)^{p}}{p!}\frac{n!}{(n-2p)!}(-2k_{0}\ell)^{n-2p}
\nonumber\\
& & \qquad\left(1-\frac{d^{2}}{\ell^{2}}\right)^{p}\left(1+\frac{d^{2}}{\ell^{2}}\right)^{p-n}
\Biggr]\,,
\label{ovl2}
\end{eqnarray}
where $[x]$ denotes the largest integer not larger than $x$, and 
the second of the above equations shows explicitly that the overlap is
well-behaved at $d=\ell$. The above expressions can be obtained by using
the explicit form of the Hermite polynomials,
\begin{eqnarray}
H_{n}(x) & = & (-1)^{n}e^{x^{2}}\frac{d^{n}}{dx^{n}}e^{-x^{2}}\\
& = & \sum_{p=0}^{\left[\frac{n}{2}\right]}\left[\frac{(-1)^{p}}{p!}\frac{n!}{(n-2p)!}
(2x)^{n-2p}\right]\,.
\end{eqnarray}
Finally, the overlap of the initial state (\ref{init}) with the
spinful eigenstates (\ref{speigen}) is given by
\begin{eqnarray}
\langle0,k,\uparrow|\psi\rangle & = & \kappa\langle0,k|\phi\rangle\,,\\
\langle n,k,\mu|\psi\rangle & = &\bar u_{n}^{\mu}\kappa\langle n,k|\phi\rangle+\bar v_{n}^{\mu}\lambda \langle n-1,k|\phi\rangle\,.
\end{eqnarray}

\subsection{Matrix elements}

As already stated, the matrix elements of the kinetic momentum as well 
as the spin operators are diagonal with respect to the
wave number $k$. For the kinetic momentum, the matrix elements read
explicitly
\begin{eqnarray}
\langle0,k,\uparrow |\pi_{x}|n,k,\mu\rangle & = & \frac{\hbar}{\ell\sqrt{2}}u_{1}^{\mu}\delta_{1,n}\,,\\
\langle0,k,\uparrow |\pi_{y}|n,k,\mu\rangle & = & \frac{-i\hbar}{\ell\sqrt{2}}u_{1}^{\mu}\delta_{1,n}\,,
\end{eqnarray}
\begin{eqnarray}
 & & \langle n_{1},k,\mu_{1}|\pi_{x}|n_{2},k,\mu_{2}\rangle\nonumber\\
 & & =\left(\sqrt{n_{1}+1}\bar u_{n_{1}}^{\mu_{1}}u_{n_{1}+1}^{\mu_{2}}
+\sqrt{n_{1}}\bar v_{n_{1}}^{\mu_{1}}v_{n_{1}+1}^{\mu_{2}}\right)\nonumber\\
 & & \qquad\times\frac{\hbar}{\ell\sqrt{2}}\delta_{n_{1},n_{2}-1}\nonumber\\
 & & +\left(\sqrt{n_{1}}\bar u_{n_{1}}^{\mu_{1}}u_{n_{1}-1}^{\mu_{2}}
+\sqrt{n_{1}-1}\bar v_{n_{1}}^{\mu_{1}}v_{n_{1}-1}^{\mu_{2}}\right)\nonumber\\
 & & \qquad\times\frac{\hbar}{\ell\sqrt{2}}\delta_{n_{1},n_{2}+1}\,,\\
 & & \langle n_{1},k,\mu_{1}|\pi_{y}|n_{2},k,\mu_{2}\rangle\nonumber\\
 & & =\left(\sqrt{n_{1}+1}\bar u_{n_{1}}^{\mu_{1}}u_{n_{1}+1}^{\mu_{2}}
+\sqrt{n_{1}}\bar v_{n_{1}}^{\mu_{1}}v_{n_{1}+1}^{\mu_{2}}\right)\nonumber\\
 & & \qquad\times\frac{-i\hbar}{\ell\sqrt{2}}\delta_{n_{1},n_{2}-1}\nonumber\\
 & & +\left(\sqrt{n_{1}}\bar u_{n_{1}}^{\mu_{1}}u_{n_{1}-1}^{\mu_{2}}
+\sqrt{n_{1}-1}\bar v_{n_{1}}^{\mu_{1}}v_{n_{1}-1}^{\mu_{2}}\right)\nonumber\\
 & & \qquad\times\frac{i\hbar}{\ell\sqrt{2}}\delta_{n_{1},n_{2}+1}\,.
\end{eqnarray}
The in-plane components of the spin operator have the matrix elements
\begin{eqnarray}
\langle0,k,\uparrow |\sigma^{x}|n,k,\mu\rangle & = & v_{1}^{\mu}\delta_{1,n}\,,\\
\langle0,k,\uparrow |\sigma^{y}|n,k,\mu\rangle & = & -iv_{1}^{\mu}\delta_{1,n}\,,
\end{eqnarray}
\begin{eqnarray}
\langle n_{1},k,\mu_{1}|\sigma^{x}|n_{2},k,\mu_{2}\rangle
 & = & \bar u_{n_{1}}^{\mu_{1}}v_{n_{1}+1}^{\mu_{2}}\delta_{n_{1},n_{2}-1}\nonumber\\
 & + & \bar v_{n_{1}}^{\mu_{1}}u_{n_{1}+1}^{\mu_{2}}\delta_{n_{1},n_{2}+1}\,,\\
 \langle n_{1},k,\mu_{1}|\sigma^{y}|n_{2},k,\mu_{2}\rangle
 & = & -i\bar u_{n_{1}}^{\mu_{1}}v_{n_{1}+1}^{\mu_{2}}\delta_{n_{1},n_{2}-1}\nonumber\\
 & - & i\bar v_{n_{1}}^{\mu_{1}}u_{n_{1}+1}^{\mu_{2}}\delta_{n_{1},n_{2}+1}\,,
\end{eqnarray}
whereas $\sigma^{z}$ is diagonal in the Landau level index $n$, and the nonvanishing
matrix elements read
\begin{eqnarray}
\langle0,k,\uparrow |\sigma^{z}|0,k,\uparrow\rangle & = & 1\,,\\
\langle n,k,\mu_{1}|\sigma^{z}|n,k,\mu_{2}\rangle & = & \bar u_{n}^{\mu_{1}}u_{n}^{\mu_{2}}-\bar v_{n}^{\mu_{1}}v_{n}^{\mu_{2}}\,.
\end{eqnarray}

\subsection{Explicit time evolution}

Using the expressions given in the previous sections, the time-evolved
expectation values of the components of the kinetic momentum can be
formulated as
\begin{widetext}
\begin{eqnarray}
\langle(\pi_{x})_{H}(t)\rangle & = & {\rm Re}\left\{\sqrt{2}\frac{\hbar}{\ell}\sum_{\mu=\pm}e^{\frac{i}{\hbar}(\varepsilon_{0}-\varepsilon_{1}^{\mu})t}
\left( |\kappa|^{2} |u_{1}^{\mu}|^{2}J_{0}+\bar \kappa\lambda u_{1}^{\mu}\bar v_{1}^{\mu}I_{0}\right)\right\}\nonumber\\
 & + & {\rm Re}\Biggl\{\sqrt{2}\frac{\hbar}{\ell}\sum_{n=1}^{\infty}\sum_{\mu_{1},\mu_{2}=\pm}
e^{\frac{i}{\hbar}(\varepsilon_{n}^{\mu_{1}}-\varepsilon_{n+1}^{\mu_{2}})t}\Biggl[
|\kappa|^{2}\left(\sqrt{n+1}|u_{n}^{\mu_{1}}|^{2}|u_{n+1}^{\mu_{2}}|^{2}
+\sqrt{n}u_{n}^{\mu_{1}}\bar v_{n}^{\mu_{1}}\bar u_{n+1}^{\mu_{2}}v_{n+1}^{\mu_{2}}\right)J_{n}\nonumber\\
 & & \qquad\qquad\qquad\qquad\qquad\qquad
+|\lambda|^{2}\left(\sqrt{n+1}\bar u_{n}^{\mu_{1}}v_{n}^{\mu_{1}}u_{n+1}^{\mu_{2}}
\bar v_{n+1}^{\mu_{2}} +\sqrt{n}|v_{n}^{\mu_{1}}|^{2}|v_{n+1}^{\mu_{2}}|^{2}\right)J_{n-1}\nonumber\\
 & & \qquad\qquad\qquad\qquad\qquad\qquad
+\bar\kappa\lambda\left(\sqrt{n+1}|u_{n}^{\mu_{1}}|^{2}u_{n+1}^{\mu_{2}}
\bar v_{n+1}^{\mu_{2}} +\sqrt{n}u_{n}^{\mu_{1}}\bar v_{n}^{\mu_{1}}|v_{n+1}^{\mu_{2}}|^{2}\right)I_{n}\nonumber\\
 & & \qquad\qquad\qquad\qquad\qquad\qquad
+\kappa\bar\lambda\left(\sqrt{n+1}\bar u_{n}^{\mu_{1}}v_{n}^{\mu_{1}}|u_{n+1}^{\mu_{2}}|^{2}
+\sqrt{n}|v_{n}^{\mu_{1}}|^{2}\bar u_{n+1}^{\mu_{2}}v_{n+1}^{\mu_{2}}\right)K_{n}\Biggr]\Biggr\}\,,
\label{px}\\
\langle(\pi_{y})_{H}(t)\rangle & = & {\rm Re}\left\{-i\sqrt{2}\frac{\hbar}{\ell}\sum_{\mu=\pm}e^{\frac{i}{\hbar}(\varepsilon_{0}-\varepsilon_{1}^{\mu})t}
\left( |\kappa|^{2} |u_{1}^{\mu}|^{2}J_{0}+\bar \kappa\lambda u_{1}^{\mu}\bar v_{1}^{\mu}I_{0}\right)\right\}\nonumber\\
 & + & {\rm Re}\Biggl\{-i\sqrt{2}\frac{\hbar}{\ell}\sum_{n=1}^{\infty}\sum_{\mu_{1},\mu_{2}=\pm}
e^{\frac{i}{\hbar}(\varepsilon_{n}^{\mu_{1}}-\varepsilon_{n+1}^{\mu_{2}})t}\Biggl[
|\kappa|^{2}\left(\sqrt{n+1}|u_{n}^{\mu_{1}}|^{2}|u_{n+1}^{\mu_{2}}|^{2}
+\sqrt{n}u_{n}^{\mu_{1}}\bar v_{n}^{\mu_{1}}\bar u_{n+1}^{\mu_{2}}v_{n+1}^{\mu_{2}}\right)J_{n}\nonumber\\
 & & \qquad\qquad\qquad\qquad\qquad\qquad
+|\lambda|^{2}\left(\sqrt{n+1}\bar u_{n}^{\mu_{1}}v_{n}^{\mu_{1}}u_{n+1}^{\mu_{2}}
\bar v_{n+1}^{\mu_{2}} +\sqrt{n}|v_{n}^{\mu_{1}}|^{2}|v_{n+1}^{\mu_{2}}|^{2}\right)J_{n-1}\nonumber\\
 & & \qquad\qquad\qquad\qquad\qquad\qquad
+\bar\kappa\lambda\left(\sqrt{n+1}|u_{n}^{\mu_{1}}|^{2}u_{n+1}^{\mu_{2}}
\bar v_{n+1}^{\mu_{2}} +\sqrt{n}u_{n}^{\mu_{1}}\bar v_{n}^{\mu_{1}}|v_{n+1}^{\mu_{2}}|^{2}\right)I_{n}\nonumber\\
 & & \qquad\qquad\qquad\qquad\qquad\qquad
+\kappa\bar\lambda\left(\sqrt{n+1}\bar u_{n}^{\mu_{1}}v_{n}^{\mu_{1}}|u_{n+1}^{\mu_{2}}|^{2}
+\sqrt{n}|v_{n}^{\mu_{1}}|^{2}\bar u_{n+1}^{\mu_{2}}v_{n+1}^{\mu_{2}}\right)K_{n}\Biggr]\Biggr\}\,,
\label{py}
\end{eqnarray}
\end{widetext}
where we have defined
\begin{eqnarray}
I_{n} & = & \int_{-\infty}^{\infty}dk\langle\phi|n,k\rangle\langle n,k|\phi\rangle\,,\\
J_{n} & = & \int_{-\infty}^{\infty}dk\langle\phi|n,k\rangle\langle n+1,k|\phi\rangle\,,\\
K_{n} & = & \int_{-\infty}^{\infty}dk\langle\phi|n-1,k\rangle\langle n+1,k|\phi\rangle\,.
\end{eqnarray}
Using Eq.~(\ref{ovl2}) it is straightforward to derive
explicit expressions for these integrals in terms of finite sums to be
evaluated numerically. As an example, for $I_{n}$ one finds
\begin{widetext}
\begin{equation}
I_{n}=\frac{\frac{d^{2}}{\ell^{2}}}{1+\frac{d^{2}}{\ell^{2}}}
\sum_{p,q=0}^{\left[\frac{n}{2}\right]}\left[\frac{(-1)^{p+q}}{p!q!}\frac{n!2^{n+1-2(p+q)}}{(n-2p)!(n-2q)!}
\left(1-\frac{d^{2}}{\ell^{2}}\right)^{p+q}
\left(\frac{1}{1+\frac{d^{2}}{\ell^{2}}}\right)^{2n-(p+q)}
R_{n-(p+q)}(\ell,d,k_{0})\right]
\end{equation}
\end{widetext}
with 
\begin{widetext}
\begin{eqnarray}
R_{m}(\ell,d,k_{0}) & = & \frac{\ell}{\sqrt{\pi}}\int_{-\infty}^{\infty}dk(k\ell)^{2m}
\exp\left(-\left(\frac{(k\ell)^{2}}{1+\frac{d^{2}}{\ell^{2}}}+d^{2}(k-k_{0})^{2}\right)\right)
\nonumber\\
 & = & \left(\frac{\ell^{4}+d^{2}\ell^{2}}{\ell^{4}+d^{2}\ell^{2}+d^{4}}\right)^{m+1/2}
\exp\left(-\frac{\ell^{4}d^{2}k_{0}^{2}}{\ell^{4}+d^{2}\ell^{2}+d^{4}}\right)
\left(-\frac{1}{4}\right)^{m}
H_{2m}\left(i\sqrt{\frac{\ell^{4}+d^{2}}{\ell^{4}+d^{2}\ell^{2}+d^{4}}}d^{2}k_{0}\right)\,.
\end{eqnarray}
\end{widetext}
Note that, despite the imaginary argument of the Hermite polynomial$H_{2m}$, 
$R_{m}$ is always real and positive. Thus computing the quantities
$I_{n}$ ( and similarly $J_{n}$ and $K_{n}$) requires again the evaluation
of Hermite polynomials. At large Landau level index $n$ this
operation limits the accuracy of the present numerical approach.
Given the data $I_{n}$, $J_{n}$, $K_{n}$, the summations (\ref{px}), (\ref{py})
are
to be performed numerically where the sum over the Landau level index $n$
can be truncated at a sufficiently large energy. For the simulations presented
in this work it is sufficient to take into account the first 25 Landau 
levels where the evaluation of Hermite polynomials is numerically
unproblematic.

It is noteworthy that the quantities $I_{n}$, $J_{n}$, $K_{n}$ 
fulfill certain sum rules which
provide a convenient check on numerical evaluations. For instance,
normalization of the initial state $|\phi\rangle$ obviously requires
\begin{equation}
\sum_{n=0}^{\infty}I_{n}=1\,.
\end {equation}
Moreover, we have
\begin{eqnarray}
\sum_{n=0}^{\infty}\sqrt{n+1}J_{n} & = & \langle\phi|a|\phi\rangle\nonumber\\
 & = & \frac{i}{\sqrt{2}}k_{0}\ell\,.
\end{eqnarray}
Analogously one derives
\begin{equation}
\sum_{n=0}^{\infty}\sqrt{n(n+1)}K_{n}=-\frac{1}{4}\frac{d^{2}}{\ell^{2}}-\frac{1}{2}k_{0}^{2}\ell^{2}\,.
\end {equation}

Finally, coming back to physical expectation values, 
the time-evolved spin components read
\begin{widetext}
\begin{eqnarray}
\langle(\sigma^{x}_{H}(t)\rangle & = & {\rm Re}\left\{2\sum_{\mu=\pm}e^{\frac{i}{\hbar}(\varepsilon_{0}-\varepsilon_{1}^{\mu})t}
\left( |\kappa|^{2}\bar u_{1}^{\mu}v_{1}^{\mu}J_{0}+\bar \kappa\lambda|v_{1}^{\mu}|^{2}I_{0}\right)\right\}\nonumber\\
 & + & {\rm Re}\Biggl\{2\sum_{n=1}^{\infty}\sum_{\mu_{1},\mu_{2}=\pm}
e^{\frac{i}{\hbar}(\varepsilon_{n}^{\mu_{1}}-\varepsilon_{n+1}^{\mu_{2}})t}\Biggl[
|\kappa|^{2}|u_{n}^{\mu_{1}}|^{2}\bar u_{n+1}^{\mu_{2}}v_{n+1}^{\mu_{2}}J_{n}
+|\lambda|^{2}\bar u_{n}^{\mu_{1}}v_{n}^{\mu_{1}}|v_{n+1}^{\mu_{2}}|^{2}J_{n-1}\nonumber\\
 & & \qquad\qquad\qquad\qquad\qquad\qquad
+\bar\kappa\lambda |u_{n}^{\mu_{1}}|^{2}|v_{n+1}^{\mu_{2}}|^{2}I_{n}
+\kappa\bar\lambda\bar u_{n}^{\mu_{1}}v_{n}^{\mu_{1}}\bar u_{n+1}^{\mu_{2}}v_{n+1}^{\mu_{2}}K_{n}\Biggr]\Biggr\}\,,\\
\langle(\sigma^{y}_{H}(t)\rangle & = & {\rm Re}\left\{-2i\sum_{\mu=\pm}e^{\frac{i}{\hbar}(\varepsilon_{0}-\varepsilon_{1}^{\mu})t}
\left( |\kappa|^{2}\bar u_{1}^{\mu}v_{1}^{\mu}J_{0}+\bar \kappa\lambda|v_{1}^{\mu}|^{2}I_{0}\right)\right\}\nonumber\\
 & + & {\rm Re}\Biggl\{-2i\sum_{n=1}^{\infty}\sum_{\mu_{1},\mu_{2}=\pm}
e^{\frac{i}{\hbar}(\varepsilon_{n}^{\mu_{1}}-\varepsilon_{n+1}^{\mu_{2}})t}\Biggl[
|\kappa|^{2}|u_{n}^{\mu_{1}}|^{2}\bar u_{n+1}^{\mu_{2}}v_{n+1}^{\mu_{2}}J_{n}
+|\lambda|^{2}\bar u_{n}^{\mu_{1}}v_{n}^{\mu_{1}}|v_{n+1}^{\mu_{2}}|^{2}J_{n-1}\nonumber\\
 & & \qquad\qquad\qquad\qquad\qquad\qquad
+\bar\kappa\lambda |u_{n}^{\mu_{1}}|^{2}|v_{n+1}^{\mu_{2}}|^{2}I_{n}
+\kappa\bar\lambda\bar u_{n}^{\mu_{1}}v_{n}^{\mu_{1}}\bar u_{n+1}^{\mu_{2}}v_{n+1}^{\mu_{2}}K_{n}\Biggr]\Biggr\}\,,\\
\langle(\sigma^{z}_{H}(t)\rangle & = & |\kappa|^{2}I_{0}+\sum_{n=1}^{\infty}\sum_{\mu_{1},\mu_{2}=\pm}
e^{\frac{i}{\hbar}(\varepsilon_{n}^{\mu_{1}}-\varepsilon_{n}^{\mu_{2}})t}\Biggl[
|\kappa|^{2}u_{n}^{\mu_{1}}\left(\bar u_{n}^{\mu_{1}}u_{n}^{\mu_{2}}-\bar v_{n}^{\mu_{1}}v_{n}^{\mu_{2}}\right)
\bar u_{n}^{\mu_{2}}I_{n}\nonumber\\
 & & \qquad\qquad\qquad\qquad\qquad\qquad
+|\lambda|^{2}v_{n}^{\mu_{1}}\left(\bar u_{n}^{\mu_{1}}u_{n}^{\mu_{2}}-\bar v_{n}^{\mu_{1}}v_{n}^{\mu_{2}}\right)
\bar v_{n}^{\mu_{2}}I_{n-1}\nonumber\\
 & & \qquad\qquad\qquad\qquad\qquad\qquad
+\bar\kappa\lambda u_{n}^{\mu_{1}}\left(\bar u_{n}^{\mu_{1}}u_{n}^{\mu_{2}}-\bar v_{n}^{\mu_{1}}v_{n}^{\mu_{2}}\right)
\bar v_{n}^{\mu_{2}}\bar J_{n-1}\nonumber\\
 & & \qquad\qquad\qquad\qquad\qquad\qquad
+\kappa\bar\lambda v_{n}^{\mu_{1}}\left(\bar u_{n}^{\mu_{1}}u_{n}^{\mu_{2}}-\bar v_{n}^{\mu_{1}}v_{n}^{\mu_{2}}\right)
\bar u_{n}^{\mu_{2}}J_{n-1}\Biggr]\,.
\end{eqnarray}
\end{widetext}


\end{document}